\documentclass[twocolumn,showpacs,preprintnumbers,amsmath,amssymb,superscriptaddress,prb]{revtex4}
\usepackage{graphicx}% Include figure files
\usepackage{dcolumn}% Align table columns on decimal point
\usepackage{bm}% bold math

\begin{document}

%\preprint{APS/123-QED}

\title{Magnetic switching of a single molecular magnet
 due to spin-polarized current}
 % Force line breaks with \\

\author{Maciej Misiorny}
\affiliation{%
Department of Physics, Adam Mickiewicz University, 61-614
Pozna\'{n}, Poland
}%
\author{J\'{o}zef Barna\'{s}}%
 \email{barnas@amu.edu.pl}
\affiliation{%
Department of Physics, Adam Mickiewicz University, 61-614
Pozna\'{n}, Poland
}%
\affiliation{%
Institute of Molecular Physics, Polish Academy of Sciences, 60-179
Pozna\'{n}, Poland
}%

\date{\today}% It is always \today, today,
             %  but any date may be explicitly specified

\begin{abstract}
Magnetic switching of a single molecular magnet (SMM) due to
spin-polarized current flowing between ferromagnetic metallic
leads (electrodes) is investigated theoretically. Magnetic moments
of the leads are assumed to be collinear and parallel to magnetic
easy axis of the molecule. Electrons tunneling through the barrier
between magnetic leads are coupled to the SMM {\it via} exchange
interaction. The current flowing through the system as well as the
spin relaxation times of the SMM are calculated from the Fermi
golden rule. It is shown that spin of the SMM can be reversed by
applying a certain voltage between the two magnetic electrodes.
Moreover, the switching may be visible in the corresponding
current-voltage characteristics.
\end{abstract}

% PACS, the Physics and Astronomy
                             % Classification Scheme.
%\keywords{Suggested keywords}%Use showkeys class option if keyword
                              %display desired
\maketitle

\section{Introduction}

Although first synthesized in the 1980s, single molecular magnets
(SMMs)~\cite{Gatteschi-SessoliAngewChem42} did not get much
attention until the beginning of the 1990s, when their unusual
magnetic properties were discovered~\cite{SessoliNature365}. Owing
to large spin and high anisotropy barrier, SMMs in a time
dependent magnetic field were shown to exhibit magnetic hysteresis
loops with characteristic steps caused by the effect of quantum
tunneling of magnetization. Current interest in SMMs is a
consequence of recent progress in nanotechnology, which enables to
attach electrodes to a single molecule and investigate its
transport properties~\cite{Heersche-etalPRL96, Jo-etalCM0603276,
Elste-TimmPRB73, Romeike-etalPRL96I, Romeike-etalPRL96II}.
Physical properties of SMMs and their nanoscale size make them a
promising  candidate for future applications in information
storage and information processing, as well as in various
spintronics devices~\cite{Timm-ElstePRB73}.

Magnetic switching of a SMM due to quantum tunneling of
magnetization in a magnetic field varying linearly in time was
considered theoretically long time ago and was also studied
experimentally~\cite{SessoliNature365}. From both practical and
fundamental reasons it would be, however, interesting to have a
possibility of switching the SMM without external magnetic field.
Such a possibility is offered by a spin polarized current. As it
is already well known, spin polarized current can switch magnetic
layers in spin valve structures, like for instance magnetic
nanopillars~\cite{Katine00}. The main objective of this paper is
just to investigate theoretically the mechanism of SMM's spin
reversal due to spin polarized current.

As a simplest system for current-induced molecular switching we
consider a SMM embedded in the barrier between ferromagnetic
electrodes (called also leads in the following). When voltage is
applied, the charge current flowing in the system is associated
with a spin current. In this paper we show that this spin current
can lead to magnetic switching of the SMM, when the voltage
surpasses a certain threshold value. Moreover, when bias increases
(linearly) in time, the switching can be observed in the
corresponding current-voltage characteristics as an additional
feature (dip or peak) in the current.

It is worth to note that spin polarized transport through
artificial quantum dots attached to ferromagnetic leads was
extensively studied in recent few years, mostly
theoretically~\cite{Rudzinski01,koenig05,Weymann06}, though some
experimental data are already available~\cite{hamaya06}. However,
investigations of spin polarized electronic transport through
molecules, and particularly through magnetic ones, are in early
stage of development.

The paper is organized as follows. In section 2 we present the
model Hamiltonian assumed to describe a molecule interacting with
magnetic leads. Theoretical analysis of electric current flowing
through the system under consideration is carried out in section
3. Numerical results on electric current and magnetic state of the
molecule are presented and discussed in section 4.

\section{Model}

\begin{figure}
\includegraphics[width=0.95\columnwidth]{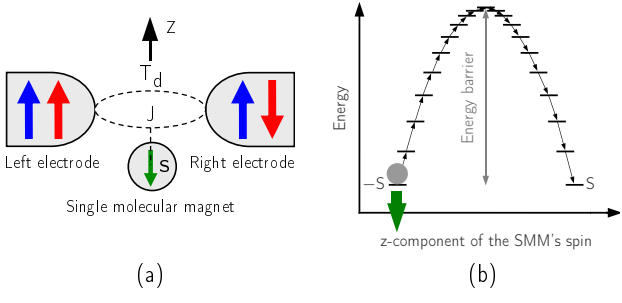}
\caption{\label{Fig1} (color online)(a) Schematic picture of the
system under consideration for two collinear configurations of the
electrodes' spin moments; parallel (blue arrows) and antiparallel
(red arrows). Dashed lines represent the two possible tunneling
processes: direct tunneling (the top line) and tunneling with
scattering on the SMM's spin due to exchange interaction (the
bottom line). (b) Energy levels corresponding to different spin
states of the SMM. The grey dot represents the initial spin state,
$|-S\rangle$, of the molecule.}
\end{figure}

We consider a model magnetic tunnel junction which consists of two
ferromagnetic leads separated by a nonmagnetic barrier, with a SMM
embedded in the barrier. Electronic transport in the system occurs
owing to tunneling processes between the leads. However, the
tunneling electrons can interact with the SMM {\it via} exchange
interaction, leading to spin switching of the molecule. For
simplicity, we will consider only collinear (parallel and
antiparallel) configurations of the leads' magnetic moments. In
addition, magnetic moments of the leads are parallel to the
magnetic easy axis of the SMM, as shown schematically in
Fig.~\ref{Fig1}(a).

For the sake of simplicity we assume that the spin number $S$ of
the molecule is constant, i.e. it does not change when current
flows through the system. This also means that the charge state of
SMM is fixed and only projection of the molecule's spin on the
quantization axis (anisotropy axis) can be changed due to the
current. In addition, we restrict the following discussion to the
case of weak coupling between the molecule and electrodes.

The full Hamiltonian of the system under consideration takes the
form
    \begin{equation}\label{eq:Ham}
    \mathcal{H}=\mathcal{H}_{S\! M\!
    M}+\mathcal{H}_L+\mathcal{H}_R+\mathcal{H}_T.
    \end{equation}
The first term describes the SMM and is assumed in the form,
    \begin{equation}\label{eq:HamSMM}
    \mathcal{H}_{S\! M\! M}=-DS_z^2,
    \end{equation}
where $S_z$ is the $z$ component of the spin operator, and $D$ is
the uniaxial anisotropy constant. Although Eq.~(\ref{eq:HamSMM})
represents the simplest Hamiltonian of a free SMM, it is
sufficient for the effects to be described here. The next two
terms describe ferromagnetic electrodes,
    \begin{equation}\label{eq:HamLeads}
    \mathcal{H}_q=\sum_{{\bf k}
    \alpha}\epsilon_{{\bf k}\alpha}^q\: a_{{\bf k}\alpha}^{q\dag}
    a_{{\bf k}\alpha}^q
    \end{equation}
for $q=L$ (left lead) and $q=R$ (right lead). The electrodes are
characterized by conduction bands with the energy dispersion
$\epsilon_{{\bf k}\alpha}^q$, where ${\bf k}$ denotes a wave
vector and $\alpha$ is the electron spin index. In Eq.(2)
$a^q_{{\bf k}\alpha}$ and $a^{q\dag}_{{\bf k}\alpha}$ are the
relevant annihilation and creation operators, respectively.

The last term of the Hamiltonian  $\mathcal{H}$ stands for the
tunneling processes~\cite{AppelbaumPRL66, AppelbaumPR67,KimPRL04},
\begin{align}
\label{eq:HamTunel}
    \mathcal{H}_T=
    \frac{1}{2}\sum_{q,q^\prime}\sum_{{\bf k}{\bf k}'\alpha\beta}
    \frac{J_{q,q^\prime}}{\sqrt{N_{q}\,N_{q^\prime}}}\:
    \bm{\sigma}_{\alpha\beta}\cdot\mathbf{S}\:
    a_{{\bf k}\alpha}^{q\dag}
    a_{{\bf k}'\beta}^{q^\prime}\
     \nonumber \\
    +\sum_{{\bf k}{\bf k}'\alpha}\frac{T_d}{\sqrt{N_L\,N_R}}\:
    a_{{\bf k}\alpha}^{L\dag}
    a_{{\bf k}'\alpha}^R\: +\: \textrm{H.c.}
    \end{align}
The first term in the above equation describes tunneling with
simultaneous interaction of tunneling electrons with the SMM {\it
via} exchange coupling, with $J_{q,q^\prime}$ being the relevant
exchange parameter. In a general case  $J_{L,L}\ne J_{R,R}\ne
J_{L,R}=J_{R,L}$. In the following, however, we assume symmetrical
situation, where $J_{L,L}=J_{R,R}=J_{L,R}=J_{R,L}\equiv J$. The
second term of Eq.~(\ref{eq:HamTunel}) describes direct tunneling
between the leads, with $T_d$ denoting the corresponding tunneling
parameter. Apart from this, $\mathbf{S}$ is the SMM's spin
operator, and $\bm{\sigma}=(\sigma^x,\sigma^y,\sigma^z)$ is the
Pauli spin operator for conduction electrons. We assume that both,
$T_d$ and $J$, are independent of energy and polarization of the
leads. Additionally, $T_d$ and $J$ are normalized in such a way
that they are independent of the size of electrodes, with $N_q$
($q=L,R$) denoting the number of elementary cells in the $q$-th
electrode.

The electric current flowing in the system is determined from the
Fermi golden rule,~\cite{KimPRL04},
    \begin{align}\label{eq:CurrentFGR}
    I&=e\sum_{mm'}\sum_{{\bf k}{\bf k}'\alpha\beta} \Bigg\{P_m\:
    W^{L{\bf k}\alpha m}_{R{\bf k}'\beta m'}\: f(\epsilon_{{\bf k}\alpha}^L)\left[1-f(\epsilon_{{\bf k}'\beta}^R)\right]
    \nonumber \\
    &\hspace{2.4cm}  - (L{\bf k}\alpha m\leftrightarrow R{\bf k}'\beta m')\Bigg\},
    \end{align}
where $e$ is the electron charge (for simplicity we assume $e>0$,
so current is positive when electrons flow from left to right),
$f(\epsilon)$ is the Fermi--Dirac distribution, $P_m$ is the
probability to find the SMM in the spin state $|m\rangle$, and
$W^{L{\bf k}\alpha m}_{R{\bf k}'\beta m'}$ is the rate of electron
transitions from the initial state $\{L{\bf k}\alpha m\}$ to the
final one $\{R{\bf k}'\beta m'\}$.

Up to the leading terms with respect to the coupling constants
$T_d$ and $J$, the current is given by the formula
    \begin{align} \label{eq:CurrentFinal}
    I& = \frac{2\pi}{\hbar}\: e^2\: \Big[|T_d|^2 + |J|^2 \left\langle S_z^2\right\rangle
    \Big] \Big(D_\uparrow^L D_\uparrow^R +D_\downarrow^L D_\downarrow^R
    \Big)\: V \allowdisplaybreaks[1]
    \nonumber\\
    & + \frac{2\pi}{\hbar}\: e\: |J|^2\: \sum_{m}  P_m
    \sum_{\eta=+,-}\eta
    \nonumber \\
    &\hspace{0.4cm} \times \Bigg\{
    D_\downarrow^L D_\uparrow^R
    A_{-\eta}(m)\zeta\Big(D(-\eta 2m+1)+ \eta eV\Big)
    \nonumber \\
    &\hspace{0.7cm}
    + D_\uparrow^L D_\downarrow^R A_\eta(m) \zeta\Big(D(\eta
    2m+1)+\eta eV\Big)
    \Bigg\}.
    \end{align}
Here, $D_\sigma^q$ is the density of states (DOS) at the Fermi
level in the $q$-th electrode for spin $\sigma$, $\left\langle
S_z^2\right\rangle=\sum_m m^2 P_m$, and $V$ is the voltage between
the leads, $eV=\mu_L-\mu_R$, with $\mu_L$ and $\mu_R$ denoting the
electrochemical potentials of the leads. Finally, $A_\pm
(m)=S(S+1)-m(m\pm1)$, and
$\zeta(\epsilon)=\epsilon\big[1-\exp(-\epsilon\beta)\big]^{-1}$
with $\beta^{-1}=k_B T$.

\section{Theoretical analysis}

To calculate electric current from Eq.~(\ref{eq:CurrentFinal}) we
need to know the probabilities $P_m$. To find them, we assume the
initial state of the SMM's spin to be $|-S\rangle$, as indicated
in Fig.~\ref{Fig1}(b). By applying a sufficiently large voltage,
one can switch the molecule to the final state $|S\rangle$. The
reversal process takes place {\it via} the consecutive
intermediate states; $|-S+1\rangle ,\ldots, |S-1\rangle$. In the
following we assume that the voltage applied to the system grows
linearly in time, $V=c\, t$, where $c$ denotes the velocity at
which the voltage is increased. This allows to observe switching
directly in the current flowing through the system when the
voltage surpasses a critical value. The probabilities $P_m$ can be
then found from the following master equations:
    \begin{equation}\label{eq:MasterEquationH=0}
    \left\{
    \begin{aligned}
    c\, \dot{P}_S\: =\: &-\: \gamma_S^- P_S\: +\: \gamma_{S-1}^+ P_{S-1},\\
    c\, \dot{P}_m\: =\: &-\: \gamma_m^- P_m\: -\: \gamma_m^+
    P_m\\
    &+\:  \gamma_{m+1}^- P_{m+1}\: +\: \gamma_{m-1}^+ P_{m-1},\\
    c\, \dot{P}_{-S}\: =\: &-\gamma_{-S}^+ P_{-S}\: +\:
    \gamma_{-S+1}^-
    P_{-S+1},
    \end{aligned}
    \right.
    \end{equation}
for $-S < m < S$ and $\dot{P}$ defined as $\dot{P}\equiv dP/dV$.
The transition rates $\gamma_m^{+(-)}$ are given by
\begin{align}\label{eq:TransitionTimes}
    \gamma_m^{+(-)}&=
    \frac{2\pi}{\hbar} |J|^2
    A_\pm(m)
    \nonumber \\
    &\hspace{0.4cm}\times \Big\{D_\uparrow^L D_\downarrow^R
    \zeta\Big(D(\pm2m+1)\pm eV\Big)
    \nonumber\\
    &\hspace{0.65cm}+D_\downarrow^L D_\uparrow^R \zeta\Big(D(\pm2m+1)\mp eV\Big)
    \nonumber \\
    &\hspace{0.65cm}+\Big[D_\uparrow^L D_\downarrow^L+D_\uparrow^R D_\downarrow^R \Big] \zeta\Big(D(\pm2m+1)\Big)
    \Big\}.
    \end{align}
The relevant boundary conditions are: $P_{-S}(V=0) = 1$ and
$P_m(V=0) = 0$, for $m\neq -S$.

\section{Numerical results and discussion}

Numerical calculations have been performed for an octanuclear
iron(III) oxo-hydroxo cluster of the formula
$\left[\textrm{Fe}_8\textrm{O}_2(\textrm{OH})_{12}(\textrm{tacn})_6\right]^{8+}$
(shortly $\textrm{Fe}_8$), whose total spin number is $S=10$. The
anizotropy constant is $D=0.292$
K~\cite{Wernsdorfer-SessoliScience284}, and we assume that $J
\approx T_d \approx 100$ meV. Furthermore, both the leads are
assumed to be made of the same metallic
 material, with the elementary cells occupied by 2 atoms
 contributing 2 electrons each. The density of free electrons is
 assumed to be $n\approx 10^{29} m^{-3}$. The electrodes are characterized by the polarization
parameter $P^q=(D_+^q\: -\: D_-^q)/(D_+^q\: +\: D_-^q)$, where
$D_{+(-)}^q$ denotes the DOS of majority (minority) electrons in
the $q$-th electrode. The temperature of the system is assumed to
be $T=0.01$ K, which is below the blocking temperature $T_B=0.36$
K of $\textrm{Fe}_8$.

    \begin{figure}
    \includegraphics[width=0.9\columnwidth]{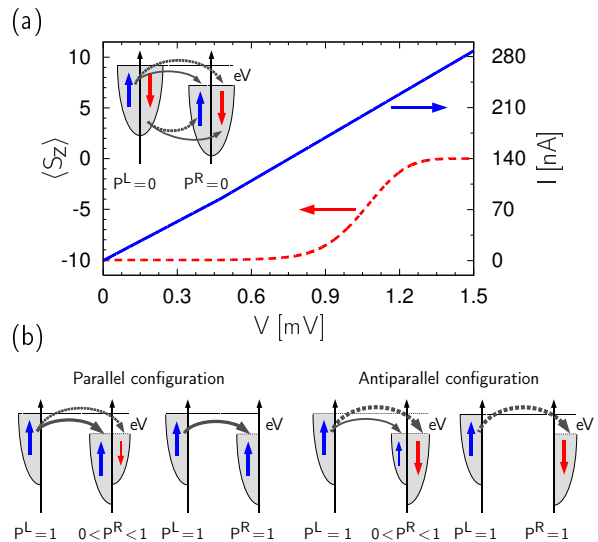}
    \caption{\label{Fig2} (color online) (a) The mean value of the SMM's spin $\left\langle
    S_z\right\rangle$ and the current $I$ flowing through the system as
    a function of voltage $V$ in the case of nonmagnetic
    electrodes, and for  $c=10$ kV/s and $T=0.01$ K.
    The inset shows schematically the density of states
    and the allowed tunneling processes (the solid arrows correspond
    to electrons tunneling without spin reversal, whereas the
    dashed arrows correspond to electrons tunneling with
    simultaneous spin-flip). (b) Schematic representation
    of the DOS and tunneling processes for both parallel and antiparallel magnetic configurations, shown in the case
    when the left electrode is made of a half-metallic ferromagnet
    while the right one is either a typical 3d ferromagnet or a half-metallic ferromagnet.
    }
    \end{figure}

Let us begin with the case where both electrodes are nonmagnetic.
In Fig.~\ref{Fig2}(a) we show the average value of the $z$
component of the SMM's spin, $\langle S_z\rangle\: =\:
\sum_{n=-S}^{S}nP_{n}$, and the charge current $I$. The spin
reversal is not found, though the current affects the SMM's spin
for $V$ exceeding the threshold voltage determined by the
anisotropy constant $D$ (energy level separation). At this
voltage, transport associated with spin-flip of the conduction
electrons becomes energetically allowed, exciting  the molecule to
the spin state $|-S+1\rangle$. As the voltage is increased
further, the different SMM's spin states $|m\rangle$ become
equally probable and $\left\langle S_z\right\rangle \to 0$.

    \begin{figure}
    \includegraphics[width=\columnwidth]{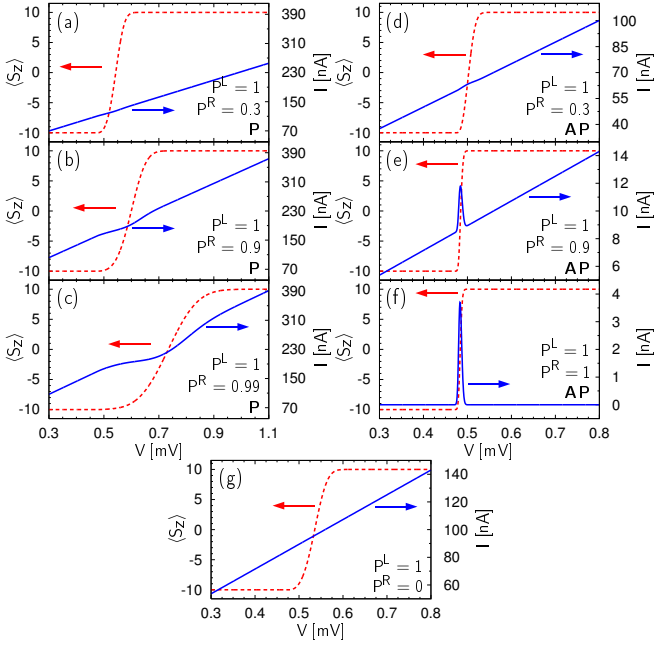}
    \caption{\label{Fig3} (color online) The mean value of the SMM's spin $\left\langle
    S_z\right\rangle$ and the current $I$ flowing through the system as
    a function of voltage V for indicated polarization parameters
    in the parallel (P) (a)--(c) and antiparallel (AP) (d)--(f)
    configurations, and for $c=10$ kV/s and $T=0.01$ K. Part  (g)
    corresponds to the case with one electrode being nonmagnetic.
    }
    \end{figure}

The situation becomes significantly different when the electrodes
are ferromagnetic, and tunneling processes are strongly spin
dependent. The following discussion is limited to the most
interesting situation, when one (say the left) electrode is a
half-metallic ferromagnet with fully spin-polarized electrons at
the Fermi level, $P^L=1$. The second electrode can be either
nonmagnetic, or typical 3d ferromagnet, or even half-metallic.
Tunneling processes with and without spin flip are indicated
schematically in Fig.2(b). The corresponding transport
characteristics and the average value of $S_z$ are shown in
Fig.~\ref{Fig3} for both parallel and antiparallel magnetic
configurations of the leads, and for various spin polarizations of
the right electrode. The complete reversal of the SMM's spin
becomes now possible, independently of the magnetic polarization
of the right electrode. Starting with the spin state $|-S\rangle$
at zero bias, one arrives at the state $|S\rangle$ when the bias
voltage surpasses the threshold value. Moreover, the switching
leads to some features in the tunneling current. We note that
switching also takes place for $0<P^L<1$, but the switching time
becomes longer.

\begin{figure}
    \includegraphics[width=\columnwidth]{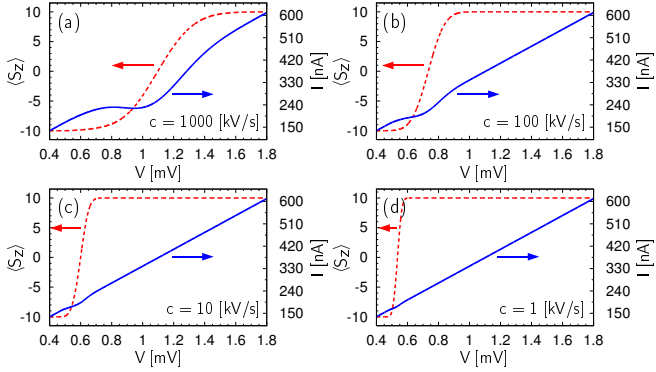}
    \caption{\label{Fig4} The mean value of the SMM's spin $\left\langle
    S_z\right\rangle$ and the current $I$ as a function of the voltage
    $V$ for various voltage sweeping speeds $c$ in the parallel magnetic
    configuration of the system. The numerical results are for $T=0.01\: K$, $P^L=1$ and $P^R=0.9$.
    }
    \end{figure}

In the parallel configuration, Figs~\ref{Fig3}(a)--(c), the
reversal process can be observed as a dip in the current, which
becomes more pronounced when $P^R\rightarrow 1$. The dip
corresponds to the voltage range where the SMM's spin reversal
process takes place. It begins at the same voltage, $V\approx
0.48$ mV, which corresponds to the energy gap between the SMM's
spin states $|-S\rangle$ and $|-S+1\rangle$ (approximately $5.55$
K in the case considered). Because the energy gaps between the
higher spin states are smaller, this energy is the activation
energy for the current induced switching. Below the threshold
voltage only direct tunneling (described by the second term in
Eq.(4)) and the non-spin-flip part of the first term in Eq.(4)
contribute to charge current. When the voltage activating spin
reversal is reached, some of the tunneling electrons can flip
their spins due to exchange interaction with the molecule, and
this leads to spin reversal of the SMM. As a result $|\langle
S_z\rangle|$ becomes reduced. This leads to partial suppression of
the non-spin-flip contribution to current from the first term in
Eq.(4). Instead of this, a spin-flip contribution becomes nonzero.
However, the latter contribution is small as it involves DOS in
the minority electron band of the right electrode, and cannot
compensate the loss of current due to the non-spin-flip tunneling
(which involves DOS for majority electrons). This leads
effectively to a dip in the current, which occurs in the voltage
range where spin switching of the SMM takes place. The dip
disappears when spin of the SMM  is completely reversed. The
broadening of the dip, in turn, stems from the fact that as $P^R
\rightarrow 1$, the transition times, $1/\gamma_m^{-(+)}$,  become
longer and longer (see Eq.~(\ref{eq:TransitionTimes})), and the
time required for complete SMM's spin reversal becomes longer as
well. This also makes the dip more pronounced.

The situation is significantly different in the antiparallel
configuration, Figs~\ref{Fig3}(d)--(f). Instead of the dip in
current, there is now a peak in the voltage range where the
switching takes place. This is because now the role of spin
minority and spin majority electron bands in the right lead is
interchanged. Additionally, the current flowing through the system
tends to 0 when $P^R\rightarrow 1$ (perfect spin valve effect),
except for a small voltage range where the reversal of the SMM's
spin occurs.

In the parallel configuration and for fully polarized electrodes
($P^L=P^R=1$), no reversal of the SMM's spin occurs and a simple
linear current-voltage characteristics is observed. On the other
hand, the linear characteristics disappears in the antiparallel
configuration, and the current does not flow through the system
except for the voltage range where the magnetic switching of the
molecule takes place, Fig.~\ref{Fig3}(f).

The probabilities $P_m$ depend on the velocity  $c$ of the voltage
increase, Eq.~(\ref{eq:MasterEquationH=0}). In Fig.4 we show
$\langle S_z\rangle$ and current $I$ in the parallel configuration
and for several values of $c$. The magnetic switching becomes
clearly visible as a dip in the current for larger values of $c$,
Fig.~\ref{Fig4}(a). At smaller values of $c$, the reversal is not
resolved in the current, Fig.~\ref{Fig4}(d). In fact, the change
in $c$ does not affect the time range within which the magnetic
switching takes place, but it only modifies the dependence between
the time and voltage scales. As a result, the transition times
$1/\gamma_m^{-(+)}$ become effectively longer within the time
scale set by the rate at which the voltage is increased. Therefore
for the higher speeds one can observe the broadening of the dip.

In summary, we showed that spin of a SMM can be reversed by a spin
polarized current, and the switching process may be visible in
current when voltage is increased in time. Full reversal of the
molecule's spin can be reached when at least one electrode is spin
polarized. The numerical results presented above apply to the case
with one electrode being fully spin polarized. However, the
current-induced switching also takes place when spin polarization
of this electrode is smaller. The switching time becomes then
appropriately longer. Moreover, for the parameters assumed in
numerical calculations the switching for positive current was only
from the state $|-S\rangle$ to $|S\rangle$. If the molecule would
be initially in the state $|S\rangle$, switching to the state
$|-S\rangle$ could be achieved by negative (reversed) current.

{\it Acknowledgements} This work is supported by funds of the
Polish Ministry of Science and Higher Education as a research
project in years 2006-2009.

\end{document}